\documentclass[aps,pra,twocolumn,showpacs]{revtex4}
\input{epsf}
\usepackage{epsfig}
\begin{document}

\title{A single cold atom as efficient stationary source of
EPR-entangled light}

\author{David Vitali} \affiliation{CNISM and Dipartimento di
Fisica, Universit\`a di Camerino, 62032 Camerino, Italy}

\author{Giovanna Morigi} \affiliation{Grup d'Optica, Departament
de Fisica, Universitat Autonoma de Barcelona, 08193 Bellaterra,
Spain}

\author{J\"urgen Eschner} \affiliation{ICFO - Institut de Ciencies Fotoniques,
Mediterranean Technology Park, 08860 Castelldefels (Barcelona),
Spain}

\date{\today}

\begin{abstract} The Stokes and anti-Stokes components of the
spectrum of resonance fluorescence of a single trapped atom, which
originate from the mechanical coupling between the scattered
photons and the quantized motion of the atomic center of mass,
exhibit quantum correlations which are of two-mode-squeezing type.
We study and demonstrate the build-up of such correlations in a
specific setup, which is experimentally accessible, and where the
atom acts as efficient and continuous source of EPR-entangled,
two-mode squeezed light. \end{abstract}


\maketitle

\section{Introduction}

The control of atom-photon interaction is object of intensive
research for its potentialities in quantum networking. In fact,
several experimental realizations have accessed novel regimes of
engineering atom-photon interactions and have opened promising
perspectives for implementing controlled nonlinear dynamics with
simple quantum optical systems. Fundamental steps in this
direction have been, amongst others, the generation of entangled
light in atomic ensembles \cite{Kimble03, Giacobino}, atomic
memory for quantum states of light \cite{LukinScience03, Lukin03,
Polzik04, KuzmichNature05, LukinNature05}, and entanglement of
remote ensembles \cite{Julsgaard01, Chou2005, Kuzmich06}. At the
single atom level, entanglement between a single atom and its
emitted photon \cite{Eberly} has been demonstrated in
\cite{Monroe04, Weinfurter2005}, while in cavity quantum
electrodynamics generation of quantum light has been achieved,
like lasing at the single atom level \cite{An94,
Kimble-atomlaser}, controlled single-photon generation
\cite{Kuhn02, Kimble-photon, Keller04, Grangier05}, as well as
quantum state and entanglement engineering in the microwave
regime~\cite{MicroCQED}.

Quantum networking with single trapped atoms or ions shows several
advantages, due to the high degree of control one can achieve on
these systems~\cite{Grangier05, Leibfried03, NeutralAtoms}.
Control can be gained on the internal as well as on the external
degrees of freedom, which can both be interfaced with light by
exchange of angular and linear momentum. In particular, by
coupling the atomic external degrees of freedom with photons via
the mechanical effect of light, atom-photon interfaces for
continuous variables can be implemented even at the level of a
single atom~\cite{Ze-Parkins99, Parkins02, PRL, Morigi06}. This
concept has been specifically applied in~\cite{PRL, Morigi06},
where the realization of a pulsed optical parametric amplifier
based on a single cold trapped atom inside a high-finesse optical
cavity was proposed, and it was shown theoretically that this
system allows for the controlled, quantum-coherent generation of
entangled light pulses by exploiting the mechanical effects of
atom-photon interaction.

In this manuscript we investigate the quantum correlations between
the Stokes and anti-Stokes sidebands of the resonance fluorescence
of a trapped atom, i.e.\ between the spectral components which are
due to the coupling of the electromagnetic field to the atom's
oscillatory motion~\cite{Lindberg86, Cirac93, Raab00, Bienert06}.
The spectrum is studied for an atom tightly confined inside a
resonator and continuously driven by a laser, in the setup
sketched in Fig.~\ref{Fig:1}. This setup has been considered
in~\cite{PRL, Morigi06} for the case of pulsed excitation, where
scattering could be considered coherent. In the present work, the
atom is continuously driven and hence both coherent and incoherent
scattering processes determine the dynamics of the system. We find
that in a suitable parameter regime the Stokes and anti-Stokes
spectral components of the resonance fluorescence are two-mode
squeezed, that is, their amplitude and phase quadratures are
quantum correlated. In fact, the variance of the difference of the
amplitude quadratures of the two sideband modes, as well as the
variance of the sum of their phase quadratures, are squeezed below
the shot noise limit, hence reproducing the salient properties of
the entangled, simultaneous eigenstate of relative distance and
total momentum of two particles, as considered in the original EPR
paradox \cite{EPR, milwal}. In our model, entanglement between the
modes originates from the mechanical coupling of the
electromagnetic field with the quantum motion of the atom, and it
is endorsed by a specific setup, which achieves resonant emission
of the Stokes and anti-Stokes photons. In this regime, the single
atom acts as an efficient continuous source of EPR-entangled,
two-mode squeezed light.

Conventionally, two-mode squeezed states emerge from the nonlinear
optical interaction of a laser with a crystal, i.e.\ from parametric
amplification or oscillation. As such, the phenomenon is the result
of many-atom dynamics, often described by a simple nonlinear
polarization model. In the single-atom case novel features appear
which are due to the coherent microscopic dynamics. Our study allows
us to identify the dependence of these features on the external
parameters, thereby giving us insight into how macroscopic
properties arise from microscopic dynamics in this particular
non-linear process. Moreover, we find peculiar spectral
characteristics of the squeezing which are unique to this system,
and which we trace back to the interplay of the various time scales
of the dynamics. In a more general context, our study is connected
to previous work on the quantum features of the spectrum of
resonance fluorescence~\cite{Aspect80, Walls81, Mandel82, Schrama92,
Nienhuis93, Narducci93, Matos94, Jakob99a, Lindberg86, Cirac93}, and
to recent experimental and theoretical studies on quantum
correlations in the light scattered by atoms~\cite{LukinScience03,
Kimble03, Polzik04, Giacobino, Raizen87, Jakob99b, Serra05}, by
semiconductor microcavities~\cite{Polaritons}, and by macroscopic
mirrors~\cite{Mancini,Pirandola03}.

This article is organized as follows. In Sec.~\ref{Sec:Review} the
basic coherent dynamics, giving rise to quantum correlations
between the Stokes and anti-Stokes components of the spectrum of
resonance fluorescence, are briefly reviewed, and the important
time scales are introduced. In Sec.~\ref{Sec:Scattering} the
theoretical model is described in detail and the relevant
scattering processes in the system are identified and discussed.
In Sec.~\ref{Sec:Spectrum} the spectrum of squeezing is evaluated
using Quantum Langevin Equations; for a quick overview of the main
results without the full theoretical elaboration, the reader may
first skip this part and jump to Sec.~\ref{Sec:Results} where the
squeezing characteristics are calculated for a specific,
experimentally achievable physical system. Finally,
Sec.~\ref{Sec:Conclusions} presents the conclusions and an
outlook.

\section{EPR-entanglement of light at the cavity output}
\label{Sec:Review}

\begin{center} \begin{figure}[htb] \epsfig{width=0.7\hsize,
file=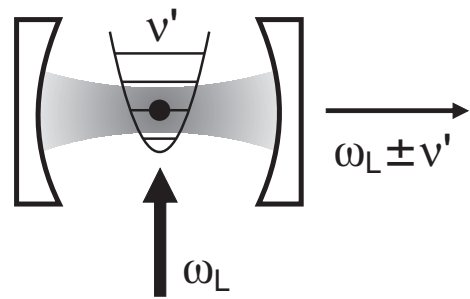} \caption{ Layout of the system. A single atom is
confined by an external potential inside an optical cavity and is
driven by a laser. The cavity is resonant with the motion-induced
Stokes and anti-Stokes components of the resonance fluorescence.
Correlations between these spectral component are measured in the
cavity output. The orientation of the considered vibrational mode
has non-zero projection onto the laser direction. A possible
geometry to implement the system would be an $F=0$ to $F'=1$
atomic transition with the quantization axis $\vec{B}$ along the
cavity axis, and $\vec{B}$, laser wave vector, and laser
polarisation mutually orthogonal, and a motional mode parallel to
the laser direction. More details can be found in
Refs.~~\cite{PRL, Morigi06} where pulsed coherent excitation was
considered. In the present paper we deal with continuous laser
excitation. } \label{Fig:1}
\end{figure} \end{center}

In this section we briefly review the coherent dynamics, described
previously in Refs.~\cite{PRL, Morigi06}, which lead to two-mode
squeezing between the Stokes and anti-Stokes modes in the light
scattered by a trapped, laser-driven atom. We thus first ignore
incoherent processes and focus on the pulsed dynamics which can be
obtained in a suitable parameter regime with a setup like the one
shown in Fig.~\ref{Fig:1}.
The trapped atom is coupled to an optical cavity of which two
modes are resonant with the Stokes and anti-Stokes sidebands,
respectively. For short times the laser-induced resonant
interaction between the center-of-mass oscillation, denoted by
annihilation and creation operators $b$ and $b^{\dagger}$, and the
two cavity modes, represented by operators $a_j$ and
$a_j^{\dagger}$ ($j=1,2$), is described by the effective
Hamiltonian in the interaction picture
\begin{eqnarray}
\label{W-eff} W_{\rm eff}={\rm i}\hbar\chi_1
a_1^{\dagger}b^{\dagger}+{\rm i}\hbar\chi_2a_2^{\dagger}b+{\rm
H.c}~, \end{eqnarray}
where the scalars $\chi_j$ indicate the strength of the coupling.
This Hamiltonian generates periodic dynamics, provided that
$|\chi_2|>|\chi_1|$, with an angular frequency
\begin{equation}
\Theta=\sqrt{|\chi_2|^2-|\chi _1|^2}~.
\label{thetabig}
\end{equation}
The time-evolution of the operators, in the Heisenberg
representation, is given by~\cite{Pirandola03}
\begin{eqnarray} \label{a_1} a_1(t)
&=&\frac{\chi_1}{\Theta}b^{\dagger}(0)\sin\Theta t +
\frac{1}{\Theta^2}\left[|\chi_2|^2-|\chi_1|^2\cos\Theta
t\right]a_1(0)~,\nonumber \\
& &-\frac{\chi_1\chi_2}{\Theta^2}\left[1-\cos\Theta t\right]a_2^{\dagger}(0)\\
a_2(t) &=&\frac{\chi_2}{\Theta}b(0)\sin\Theta t+
\frac{\chi_1\chi_2}{\Theta^2}\left[1-\cos\Theta t\right]a_1^{\dagger}(0)\nonumber\\
& &-\frac{1}{\Theta^2}\left[|\chi_1|^2-|\chi_2|^2\cos\Theta t\right]a_2(0)~,\\
b(t) &=&b(0)\cos\Theta
t+\frac{1}{\Theta}\left[-\chi_2^*a_2(0)+\chi_1 a_1^{\dagger}(0
)\right]\sin\Theta t~.\nonumber\\
\label{b}
\end{eqnarray}
In general these solutions describe tripartite entanglement among
cavity modes and center-of-mass oscillator~\cite{Pirandola03}. An
interesting situation is found after half a period, for
$T_{\pi}=\pi/\Theta$. At this time (modulus $2\pi$) the
center-of-mass oscillator is uncorrelated with the cavity modes,
which exhibit EPR-type entanglement~\cite{PRL,Morigi06}.

Clearly, this description is approximate, and valid only when
incoherent processes can be neglected. In the present work we
consider the situation in which the atom is continuously driven by
the laser field, such that quantum noise and dissipative processes
affect the dynamics relevantly. We show that steady state
entanglement, i.e.\ quantum-correlated spectral fluctuations in
the two-mode cavity output field, is found also under these
conditions. The details of this entanglement will depend on the
comparison between the time scale set by the coherent dynamics,
$\Theta^{-1}$, and the time scales of the dissipative processes,
$\kappa^{-1}$ for loss of photons from the cavity, and
$\gamma^{-1}$ for spontaneous scattering from the atom. In
particular, we will show that the squeezing spectrum shows
distinct, qualitatively different features in the regimes
$\Theta<\kappa$, $\Theta=\kappa$, and $\Theta>\kappa$. The reader
is referred to Sec.~\ref{Sec:Results}, where the spectra for
different parameter regimes are reported.

\section{Scattering processes}
\label{Sec:Scattering}

The purpose of this section is to discuss the coherent and
incoherent scattering processes determining the dynamics of the
system. We will present these processes using physical pictures
derived from the scattering matrix under moderate simplifications,
in order to illustrate the more rigorous derivations presented in
the subsequent section. We first introduce the model, and then
identify the scattering processes and determine the corresponding
rates.

\subsection{Model} \label{Sec:II}

We consider an atom of mass $M$ inside an optical resonator and
driven by a laser. The atomic motion is confined by an external
potential, which we assume sufficiently steep in the radial
direction so that the motion in this plane can be considered
frozen out. We denote by $x$ the axis of the remaining
one-dimensional atomic center-of-mass motion. Moreover, we assume
that only the atomic dipole transition between ground state
$|g\rangle$ and excited state $|e\rangle$ couples relevantly to
the fields, such that we can restrict the electronic dynamics to
these two states. The atomic dipole is laser-driven, and it
couples to two modes ($j=1,2$) of the resonator, as well as to the
external modes of the electromagnetic field. The cavity modes
couple also to the external modes of the electromagnetic field
through the imperfect mirrors of the resonator. The total dynamics
is governed by the Hamiltonian
$$H=H_0+W,$$
where $H_0$ is the self-energy of the system of atom and fields, and
$W$ describes their mutual interaction, as well as the coupling
between the cavity modes and the external modes through the finite
transmission at the cavity mirrors. We now introduce each term in
detail, and discuss the dynamics in the reference frame of the laser
at the angular frequency $\omega_L$. We decompose $H_0$ according to
\begin{equation} H_0=H_a+H_c+H_{\rm emf}~. \end{equation}
Here, $H_a$ is the Hamiltonian for the relevant atomic degrees of
freedom, \begin{equation} H_a=-\hbar\Delta|e\rangle\langle e|+H_{\rm
mec}~, \label{H:atom} \end{equation} where
$\Delta=\omega_L-\omega_0$ is the detuning of the laser from the
dipole transition at the angular frequency $\omega_0$, and
\begin{equation}
\label{H:mec} H_{\rm
mec}=\hbar\nu\left(b^{\dagger}b+\frac{1}{2}\right) \end{equation}
describes the harmonic motion of the atomic center of mass at
angular frequency $\nu$, as determined by an external potential,
where $b,b^{\dagger}$ are the annihilation and creation operators,
respectively, of a quantum of vibrational energy $\hbar\nu$. In
particular, the atomic position is given by
$x=\sqrt{\hbar/2M\nu}(b+b^{\dagger})$.  We denote by $|n\rangle$ the
eigenstates of $H_{\rm mec}$ at energy $\hbar\nu (n+1/2)$. The
Hamiltonian for the cavity modes, which couple appreciably to the
dipole transition, is
\begin{equation} H_c=-\sum_{j=1,2}\hbar\delta_ja_j^{\dagger}a_j~,
\end{equation}
where $\delta_j=\omega_L-\omega_j$ are the detunings of the laser
from the frequencies $\omega_j$ of two optical modes, and
$a_j,a_j^{\dagger}$ are the respective annihilation and creation
operators of a quantum of energy $\hbar\omega_j$, i.e.\ a photon
in mode $j$. We denote by $|n_1,n_2\rangle$ the eigenstates of
$H_c$ at energy $-\hbar\delta_1n_1-\hbar\delta_2n_2$, and consider
the situation in which the mode frequencies fulfill the relation
\begin{equation}
\omega_2 - \omega_1 = 2\nu^{\prime}~.
\end{equation}
where
\begin{equation}
\nu^{\prime}=\nu+\delta\nu
\end{equation}
and $\delta\nu$ takes into account radiative shifts, such that cavity modes 1 and 2 can be simultaneously resonant
with the Stokes and the anti-Stokes transitions. This contribution will be discussed in Sec.~\ref{Sec:ac-Shift} and determined in Sec.~\ref{Sec:QLE}.

Finally, the modes of the electromagnetic field external to the
cavity possess the free Hamiltonian
$$H_{\rm emf}=-\hbar\sum_{\bf
k_j}\delta_{\bf k_j}r^{\dagger}_{\bf k_j}r_{\bf k_j}
-\hbar\sum_{\bf k_s}\delta_{\bf k_s}r^{\dagger}_{\bf k_s}r_{\bf
k_s}~,$$
where $r_{\lambda}$, $r^{\dagger}_{\lambda}$ are annihilation and
creation operators, respectively, of a photon at angular frequency
$\omega_{\lambda}=\omega_L-\delta_{\lambda}$, wavevector ${\bf
k_{\lambda}}$ and polarization ${\bf e_{\lambda}}$. Here, the
subscripts $\lambda={\bf k_s}$ and $\lambda={\bf k_j}$ indicate the
modes of the field which couple to the dipole and to the cavity
modes (through the mirrors), respectively. The interaction term
\begin{equation}
W=H_{aL}+H_{ac}+W_{\bf k_s}+W_{\bf k_j}
\end{equation}
describes the couplings among atom and fields, decomposed into four
terms which correspond to the coupling between atom and laser
($H_{aL}$), atom and cavity modes ($H_{ac}$), atom and modes of the
external electromagnetic field ($W_{\bf k_s}$), and cavity modes and
external electromagnetic field ($W_{\bf k_j}$). We discuss these
terms in the Lamb-Dicke regime, when the atomic motion is well
localized over the wavelengths of the fields, such that the
Lamb-Dicke parameter $\eta=\sqrt{\hbar k^2/2M\nu}$ is small,
$\eta\ll 1$.
At lowest order in $\eta$, the coupling between laser
and dipole has the form~\cite{FootnoteEta}
\begin{eqnarray}
\label{HL} H_{aL} &=&\hbar\Omega\sigma^{\dagger}
\Bigl[\left(1-\frac{\eta^2}{2}\cos^2\theta_L(2b^{\dagger}b+1)\right)\\
&+&{\rm i}\eta\cos\theta_L(b^{\dagger}+b) +{\rm O}(\eta^2)\Bigr]
+{\rm H.c.}~,\nonumber
\end{eqnarray}
with $\sigma=|g\rangle\langle e|$ the dipole lowering operator and
$\sigma^{\dagger}$ its adjoint, $\Omega$ the Rabi frequency, and
$\theta_L$ the angle between the direction of propagation of the
laser and the motional axis $\hat{x}$. In what follows we denote
the moduli of all relevant wave vectors by $k$, as their
differences are negligible. The coupling between the dipole and
the cavity modes is represented by
\begin{eqnarray} \label{Hint} H_{ac} &=& \hbar\sum_{j=1,2}
g_j\cos\phi_ja_j\sigma^{\dagger}
\Bigl[\left(1-\frac{\eta^2}{2}\cos^2\theta_c(2b^{\dagger}b+1)\right) \nonumber \\
&-&\eta\cos\theta_c\tan\phi_j(b^{\dagger}+b)\Bigr]+{\rm H.c. +{\rm
O}(\eta^2)}~,
\end{eqnarray}
where $g_j$ is the coupling strength of the dipole to mode $j$,
and the cavity axis forms an angle $\theta_c$ with the axis
$\hat{x}$ of the motion. The angle $\phi_j$ takes into account the
position of the trap center inside the standing wave of the
cavity. Finally, the terms
\begin{eqnarray*}
W_{\bf k_s} &=&\sum_{\bf k_s}\hbar g_{\bf
k_s}\sigma^{\dagger}r_{\bf k_s}\Bigl[\left(1-
\frac{\eta^2}{2}\cos^2\theta_{\bf k_s}(2b^{\dagger}b+1)\right)\\
&+&{\rm i}\eta\cos\theta_{\bf k_s}(b+b^{\dagger})+{\rm O}(\eta^2)\Bigr]+{\rm H.c.})~,\nonumber\\
W_{\bf k_j} &=&\sum_{\bf k_j}\hbar g_{\bf k_j}(a_j^{\dagger}r_{\bf
k_j}+{\rm H.c.})
\end{eqnarray*}
describe the coupling of atom and cavity to the modes of the
external e.m.-field. Here, $W_{\bf k_s}$ is the coupling of the
dipole, at Rabi frequencies $g_{\bf k_s}$, with the external modes,
whose wave vectors form angles $\theta_{\bf k_s}$ with the motional
axis. This coupling gives rise to the finite linewidth $\gamma$ of
the excited state, $\gamma=2\pi\rho_{\bf k_s}(\omega_0)|g_{\bf
k_s}(\omega_0)|^2$, with $\rho_{\bf k_s}(\omega_0)$ density of
states of the e.m.-field coupling to the atomic dipole at angular
frequency $\omega_0$. The term $W_{\bf k_j}$ describes the coupling
of the cavity modes with the external modes at strength $g_{\bf
k_j}$. This coupling gives rise to the linewidth of the cavity modes
$\kappa_j=\pi |g_{\bf k_j}|^2\rho_{\bf k_j}(\omega_j)$, with
$\rho_{\bf k_j}(\omega_j)$ density of states of the e.m.-field
coupling to the cavity modes at angular frequency $\omega_j$.

\subsection{Basic scattering processes}
\label{Sec:Scattering:I}

We consider the limit in which the atom is far-detuned from cavity
modes and laser, $|\Delta| \gg \gamma, \delta_j, g_j, \Omega$. In
this limit all terms of $W$ are weak perturbations to the
dynamics. We assume that the system is in the initial state
\begin{equation} |\psi_i\rangle=|g,n;0_1,0_2;0_{\bf k_j};0_{\bf
k_s}\rangle,
\end{equation}
with energy $E_i=\hbar\nu n$, where the atom is in the ground
state $|g\rangle$, the center-of-mass oscillator is in the number
state $|n\rangle$, and the cavity modes and the external
e.m.-field are in the vacuum state, $|0_1,0_2;0_{\bf k_j};0_{\bf
k_s}\rangle$. The scattering matrix elements between the initial
state and all possible final states $|\psi_f\rangle$, with energy
$E_f$, have the form
\begin{equation} {\cal S}_{if}=\delta_{if}-2\pi{\rm
i}\delta(E_f-E_i){\cal T}_{if} \end{equation}
where $\delta_{if}$ is the Kronecker-delta, $\delta(E_f-E_i)$ is a
delta-function giving energy conservation between initial and
final states, and ${\cal T}_{if}$ is the transition matrix to be
evaluated in lowest order in perturbation theory,
\begin{eqnarray*} {\cal T}_{if}=\langle
\psi_f|W|\psi_i\rangle+\langle \psi_f|W\frac{1}{E_i-H_{\rm
eff}}W|\psi_i\rangle
\end{eqnarray*}
with
\begin{equation} H_{\rm
eff}=-\hbar\left(\Delta+{\rm i}\frac{\gamma}{2}\right)
|e\rangle\langle e|+\hbar\nu b^{\dagger}b
-\hbar\sum_{j=1,2}\left(\delta_j+{\rm
i}\kappa_j\right)a_j^{\dagger}a_j \end{equation}
We now consider all possible scattering transitions to resonant
states, i.e.\ to final states $|\psi_f\rangle$ at energy
$E_f=E_i$.
\begin{center}
\begin{figure*}[htb]
\epsfig{width=0.8\textwidth, file=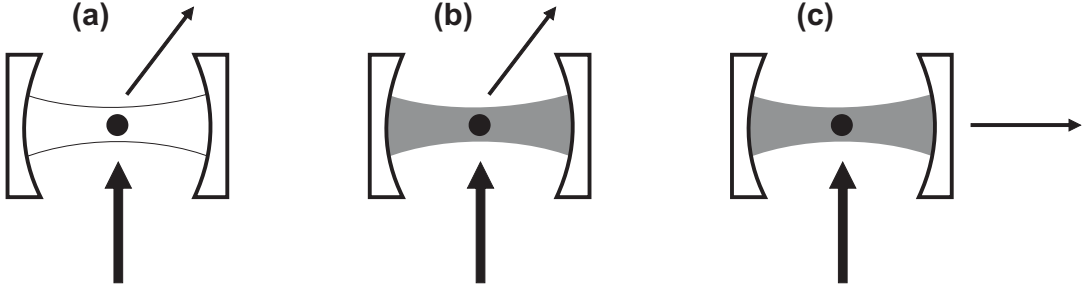} %
\caption{Basic scattering processes. (a): A laser photon is
absorbed and emitted by the atom, without coupling to the cavity
mode. (b) A laser photon is scattered into the cavity mode and
then rescattered by the atom into the external modes of the
electromagnetic field. (c) A laser photon is scattered by the atom
into the cavity mode, and then it is transmitted by the cavity
mirror into the modes of the external electromagnetic field.}
\label{Fig:0b}
\end{figure*}
\end{center}

\subsubsection{Scattering of laser photons into the external
e.m.-field}
\label{Laser:Free}

We consider the scattering of a laser photon into the external
e.m.-field by spontaneous emission, hence coupling of
$|\psi_i\rangle$ to the final states $|\psi_{\bf
k_s}\rangle=|g,n^{\prime};0_1,0_2;0_{\bf k_j};1_{\bf k_s}\rangle$.
This process is sketched in Fig.~\ref{Fig:0b}(a). Here, the
coupling with the cavity mode is neglected, as the cavity is
far-detuned from the dipole, and the rate of this process can be approximated by
the scattering rate of the atom in free space,
\begin{eqnarray} \Gamma_{if}^{\rm sp}
&\approx&\frac{1}{\gamma}\Bigl(|t_{0}^{\rm
sp}|^2\delta_{n^{\prime},n}+|t_{+1}^{\rm
sp}|^2(n+1)~\delta_{n^{\prime},n+1}\\ & & +|t_{-1}^{\rm
sp}|^2n~\delta_{n^{\prime},n-1}\Bigr), \nonumber
\end{eqnarray}
where
\begin{eqnarray}
&&t_{0}^{\rm sp}=\frac{\gamma\Omega}{\Delta+{\rm i}\gamma/2},\label{T:sp:0}\\
&&t_{+1}^{\rm
sp}=\eta\gamma\Omega\left(\frac{\cos\theta_L}{\Delta-\nu+{\rm
i}\gamma/2}+\frac{\cos
\theta_{\bf k_s}}{\Delta+{\rm i}\gamma/2}\right)\label{T:sp:+},\\
&&t_{-1}^{\rm
sp}=\eta\gamma\Omega\left(\frac{\cos\theta_L}{\Delta+\nu+{\rm
i}\gamma/2}+\frac{\cos\theta_{\bf k_s}}{\Delta+{\rm
i}\gamma/2}\right)~. \label{T:sp:-}
\end{eqnarray}
The process described by amplitude~(\ref{T:sp:0}) does not affect
the dynamics of the cavity modes nor that of the center-of-mass
motion. In contrast, the amplitudes~(\ref{T:sp:+})
and~(\ref{T:sp:-}) are coherent superpositions of scattering
processes involving, respectively, the mechanical effect of the
laser and of the emitted photon on the atomic
motion~\cite{Cirac93, Bienert06}, thereby affecting the coherence
of the motional state. Their rate is $\gamma_b \approx \eta^2
(\cos^2\theta_L + \alpha) \gamma \Omega^2/ \Delta^2$, where
$\alpha$ describes the angular dispersion of the spontaneously
emitted photons, determined by the quantum numbers of the atomic
transition \cite{Stenholm86}.

\subsubsection{Scattering of laser photons into the cavity modes}
\label{Sec:Chi}

Next we discuss the processes in which a laser photon is scattered
into one of the cavity modes, thereby coupling the initial state
$|\psi_i\rangle$ to the states
$|\psi_1\rangle=|g,n^{\prime};1_1,0_2;0_{\bf k_j};0_{\bf
k_s}\rangle$ or $|\psi_2\rangle=|g,n^{\prime};0_1,1_2;0_{\bf
k_j};0_{\bf k_s}\rangle$.
As these states are not stable, but resonantly coupled to the
continuum of states $|g,n^{\prime}; 0_1,0_2; 1_{\bf k_j}; 0_{\bf
k_s}\rangle$ by cavity decay, the correct final states of these
scattering processes describe the processes sketched in
Fig.~\ref{Fig:0b}(c) and have the form
\begin{equation} |\psi_{\bf k_j}\rangle=\sqrt{Z_{\bf
k_j}}\left(1+\frac{Q_j}{E_{\bf k_j}-H}W_{\bf
k_j}\right)|\psi_j\rangle,
\end{equation}
where $Q_j$ projects onto the subspace orthogonal to
$|\psi_j\rangle$, and $Z_{\bf k_j}$ ensures the normalization of
the state. Furthermore, $Z_{\bf k_j}$ gives the occupation
probability of state $|\psi_j\rangle$, since $Z_{\bf
k_j}=|\langle\psi_j|\psi_{\bf k_j}\rangle|^2$.

The coupling rate between state $|\psi_i\rangle$ and states
$|\psi_{\bf k_j}\rangle$ takes the form
\begin{eqnarray}
\Gamma_{if_j}^{\rm cav}
&\approx&\frac{2\kappa_j}{\delta_j^2+\kappa_j^2}~|t_{0}^{\rm cav}|^2~\delta_{n^{\prime},n}\label{sum}\\
& &+\frac{2\kappa_j}{(\delta_j-\nu)^2+\kappa_j^2}~
|t_{j,+}^{\rm cav}|^2(n+1)~\delta_{n^{\prime},n+1}\nonumber\\
& &+\frac{2\kappa_j}{(\delta_j+\nu)^2+\kappa_j^2}~|t_{j,-}^{\rm
cav}|^2n~\delta_{n^{\prime},n-1}\nonumber,
\end{eqnarray}
where
\begin{eqnarray}
&&t_{0}^{\rm cav}=\Omega g_j^{*}\cos\phi_j\frac{1}{\Delta+{\rm i}\gamma/2},\\
&&t_{j,+}^{\rm cav}=\eta \Omega g_j^{*}\cos\phi_j\left[\frac{{\rm
i}\cos\theta_L}{\Delta-\nu+{\rm i}\gamma/2}
-\frac{\cos\theta_c\tan\phi_j}{\Delta+{\rm i}\gamma/2}\right],\label{T:int:1}\\
&&t_{j,-}^{\rm cav}=\eta \Omega g_j^{*}\cos\phi_j\left[\frac{{\rm
i}\cos\theta_L}{\Delta+\nu+{\rm
i}\gamma/2}-\frac{\cos\theta_c\tan\phi_j}{\Delta+{\rm
i}\gamma/2}\right].\label{T:int:2}\end{eqnarray}
Like in Eqs.~(\ref{T:sp:+}) and (\ref{T:sp:-}), we recognize on
the RHS of Eqs.~(\ref{T:int:1}) and~(\ref{T:int:2}) the coherent
addition of two scattering amplitudes, here representing the
mechanical effects of the laser and of the cavity,
respectively~\cite{Cirac95, Zippilli05}. These processes are at
the basis of the coherent coupling between the atomic motion and
the cavity modes described by Hamiltonian~(\ref{W-eff}), where
$\chi_1=-{\rm i}t_{1,+}^{\rm cav}$ and $\chi_2=-{\rm
i}t_{2,-}^{\rm cav}$. We are interested in the regime where energy
can be stored in the cavity modes through this coupling, which
requires $|\chi_1|,|\chi_2|\gg \gamma_b$ as a necessary condition.
In this situation, it is visible from the equations that in the
limit $\kappa_j\ll\nu$, by choosing $\delta_1=\nu$ and
$\delta_2=-\nu$ one can achieve the optimum enhancement of the
scattering of a laser photon into mode 1 accompanied by the
excitation of the motion by one vibrational quantum, and of the
scattering of a laser photon into mode 2 accompanied by the
de-excitation of the motion by one vibrational quantum. Note that
these scattering terms, $t_{j,\pm}^{\rm cav}$, have an incoherent
component which scales with $\gamma/|\Delta\pm\nu|$. Therefore, in
general coherent dynamics can only be achieved when
$\gamma\ll|\Delta|$, on a time scale such that incoherent terms
are negligible. Moreover, the condition $\gamma\ll\nu$ is also
required in order to create quantum correlations between the two
cavity modes, since the difference between the two coupling
strengths $\chi_1$ and $\chi_2$ determines the typical time scale
on which entanglement is established, see Sec.~\ref{Sec:Review}
and~\cite{Morigi06}.

\subsubsection{Scattering of cavity photons into the external e.m.-field}
\label{Sec:cav-noise}

Assuming that photons have been coherently scattered into the
cavity modes, they can be re-absorbed by the atom and emitted
spontaneously into the external e.m.-field, as sketched in
Fig.~\ref{Fig:0b}(b). In order to focus on the evaluation of the
corresponding element of the scattering matrix, we consider the
regime of very small cavity loss rate, i.e.\ we assume stable
cavity modes and ignore, for the clarity of the picture, cavity
decay. Be the initial state
\begin{equation}
\label{Initial:2} |\psi_{i,m}\rangle=|g,n;m_1,m_2;0_{\bf
k_j};0_{\bf k_s}\rangle
\end{equation}
at energy $E_{i,m} = \hbar\nu n - \hbar m_1 \delta_1 - \hbar m_2
\delta_2$, with the atom in $|g\rangle$, the center-of-mass
oscillator in the number state $|n\rangle$, the cavity modes in
the Fock states $|m_1\rangle$ and $|m_2\rangle$, and the external
e.m.-field in the vacuum state, $|0_{\bf k_j};0_{\bf k_s}\rangle$.
This state is coupled to the states
\begin{eqnarray}
|\psi_{f,m_1^{\prime}}\rangle=|g,n;m_1-1,m_2;0_{\bf k_j};1_{\bf k_s}\rangle\\
|\psi_{f,m_2^{\prime}}\rangle=|g,n;m_1,m_2-1;0_{\bf k_j};1_{\bf
k_s}\rangle
\end{eqnarray}
by absorption of a cavity photon and spontaneous emission. We
evaluate the corresponding rate under the assumption, that
$\tan\phi_j=0$, i.e., there are no mechanical effects of the
resonator on the atom at first order in $\eta$, and find an
effective loss rate of the cavity modes
\begin{eqnarray} \Gamma_{if_j}^{\rm cav-sp}
&=&\gamma|g_j|^2\left|\frac{\sqrt{m_j}}{\Delta-\delta_j+{\rm
i}\gamma/2}\right|^2~. \label{cav-sp}
\end{eqnarray}

It should be noted that these processes arise from atomic
scattering of a laser photon into the cavity modes, which is then
rescattered by atomic emission into the external modes of the
e.m.-field. Hence, these processes can interfere with atomic
scattering of a laser photon, in the limit discussed in
Sec.~\ref{Laser:Free}, in which the coupling to the cavity plays
no role. In these calculations we have not considered the coherent
addition of these two noise effects, but we will consider phase
relations and possible interference in these noise sources when
studying the dynamics with the quantum Langevin equations in
Sec.~\ref{Sec:QLE}.

\subsubsection{a.c.-Stark shift of the ground state energy}
\label{Sec:ac-Shift}

Since the efficiency of production of two-mode squeezed light is
based on the resonant enhancement of two-photon processes, it is
important to consider systematically radiative corrections to the
resonance frequencies in the implementation of the dynamics in
Sec.~\ref{Sec:Review}. Therefore, we now evaluate corrections to the
energy of state $|\psi_{i,m}\rangle$, Eq.~(\ref{Initial:2}), due to
far-off resonance coupling in the limit of very small cavity decay
rates. When considering the a.c.-Stark shift of state
$|\psi_{i,m}\rangle$, we find three contributions, each associated
to a different kind of coupling: (i) the a.c.-Stark shift due to the
off-resonant laser coupling with the excited state at zero order in
the mechanical effects, $\delta\omega_0\sim\Omega^2/\Delta$ for
$|\Delta|\gg\gamma$. It leads to a shift $\delta\omega_0$ of the
dipole resonance frequency. The mechanical effects of the laser on
the atoms give rise to (ii) a contribution which is linear in the
number of vibrational excitation, and can hence be considered a
renormalization of the trap frequency. This a.c.-Stark shift reads
\begin{eqnarray} \label{deltanu_b} \delta\nu_b
&\approx&\eta^2\cos^2\theta_L\Omega^2 b^{\dagger}b \\
&\times &{\rm Re}\left\{
\frac{1}{\Delta+\nu+{\rm i}\gamma/2}+\frac{1}{\Delta-\nu+{\rm i}\gamma/2}-\frac{1}{\Delta+{\rm i}\gamma/2}\right\}\nonumber \\
&=      &\eta^2\cos^2\theta_L\Omega^2~b^{\dagger}b \nonumber\\
&\times
&\left(\frac{2\Delta(\Delta^2-\nu^2+\gamma^2/4)}{(\Delta^2-\nu^2+\gamma^2/4)^2+\gamma^2\nu^2}
-\frac{\Delta}{\Delta^2+\gamma^2/4}\right).\nonumber
\end{eqnarray} Finally, off-resonant coupling of the cavity mode
with the dipole transition gives rise to an a.c.-Stark shift of
the cavity mode levels, which reads at leading order
\begin{equation} \label{deltaomega_j}
\delta\omega_j\approx\frac{|g_j|^2\cos^2\phi_j
(\Delta-\delta_j)}{(\Delta-\delta_j)^2+\gamma^2/4}a_j^{\dagger}a_j+{\rm
O}(\eta^2). \end{equation} These shifts are in general not small
and should be taken into account, when aiming at the resonant
enhancement of certain processes over others. It should be
remarked that the correction to $\delta\omega_j$ in
Eq.~(\ref{deltaomega_j}) which is at second order in $\eta$ arises
from the mechanical effects of the interaction between resonator
and center-of-mass motion. This term is non-linear, as it is a
shift which depends on the number of vibrational excitation, but
is a negligible contribution to $\delta\omega_j$. On the other
hand, this term gives rise to an additional contribution to the
a.c.-Stark shift of the center-of-mass motion, which is of the
same order as $\delta\nu_b$ and depends on the number of photons.
Its effect is detrimental, as the resulting spectrum of the
center-of-mass excitations deviates from the one of a harmonic
oscillator. In the system we consider we will neglect this
contribution, focussing onto the regime in which the mechanical
effects of the cavity mode can be neglected. This corresponds to
situations, where the motion, for instance, is almost orthogonal
to the cavity wave vector, $|\cos\theta_c|\ll 1$.

\section{Spectrum of light at the cavity output}
\label{Sec:Spectrum}

In this section we evaluate the spectrum of the light transmitted
by the cavity mirror. The spectrum is best evaluated using the
quantum Langevin equations for the operators $a_j$,
$a_j^{\dagger}$ and $b$. The equations we obtain are rather
involved, however the physical meaning of each term can be
identified by comparison with the rates of the scattering
processes discussed in the previous section.

\subsection{Quantum Langevin Equations} \label{Sec:QLE}

We shall study the dynamics using the quantum Langevin equations
(QLE) of the system. For convenience, we write the interaction
Hamiltonian of the atom with the laser and the cavity fields as
\begin{equation}\label{hinteff}
  H_{int}=H_{aL}+H_{ac}=\hbar\left(\sigma^{\dagger}B+\sigma B^{\dagger}\right),
\end{equation}
where
\begin{eqnarray}
  && B=\Omega\left(1-\frac{\eta^2}{2}\cos^2 \theta_L (2 b^{\dagger}b+1)\right)+i\eta\Omega\cos\theta_L (b^{\dagger}+b) \nonumber
  \\
  &&+\sum_{j=1,2}g_j \cos\phi_j a_j\left(1-\frac{\eta^2}{2}\cos^2 \theta_c (2
  b^{\dagger}b+1)\right)\nonumber\\
  &&-\eta \cos\theta_c g_j \sin\phi_j a_j (b^{\dagger}+b),
  \label{beff}
\end{eqnarray}
The QLE read \begin{eqnarray}
\dot{a}_1(t)&=&i\delta_1 a_1(t)+i\sigma(t)\left[B(t)^{\dagger}, a_1(t)\right] \nonumber \\
&& -\kappa_1 a_1(t)+\sqrt{2\kappa_1}a_1^{in}(t), \\
\dot{a}_2(t)&=&i\delta_2 a_1(t)+i\sigma(t)\left[B(t)^{\dagger}, a_2(t)\right] \nonumber \\
&& -\kappa_2 a_2(t)+\sqrt{2\kappa_2}a_2^{in}(t), \\
\dot{b}(t)&=&-i\nu b(t)+i\sigma(t)\left[B(t)^{\dagger},
b(t)\right]+
i\sigma(t)^{\dagger}\left[B(t), b(t)\right] \nonumber \\
&& -\kappa_b b(t)+\sqrt{2\kappa_b}b^{in}(t), \\
\dot{\sigma}(t)&=&\left[i\Delta -\frac{\gamma}{2}\right] \sigma(t)+\sigma_z(t) \left( i B(t)+\sqrt{\gamma}f^{in}(t)\right), \label{penlastqle}\\
\dot{\sigma}_z(t)&=& 2i\sigma(t) B(t)^{\dagger}-2i\sigma^{\dagger}(t) B(t)-\gamma \left[\sigma_z(t)+1\right] \nonumber \\
&&-2\sigma^{\dagger}(t)\sqrt{\gamma}f^{in}(t)-2\sigma(t)\sqrt{\gamma}f^{in}(t)^{\dagger},
\label{lastqle} \end{eqnarray} where
$\sigma_z=\sigma^{\dagger}\sigma-\sigma \sigma^{\dagger}$, and we
have introduced the vacuum input noises $a_j^{in}(t)$ ($j=1,2$) of
the cavity modes with corresponding decay rate $\kappa_j$, the
spontaneous emission noise $f^{in}(t)$ at rate $\gamma$, and we
also added a phenomenological input noise $b^{in}(t)$ acting on
the atom's motion, describing the heating at rate $\kappa_b$ due
to the fluctuations of the trap potential. These four noise
sources are mutually uncorrelated and have zero mean, while their
second-order correlations have the form \begin{eqnarray}
&&\langle a_1^{in}(t)a_1^{in}(t')^{\dagger}\rangle = \langle a_2^{in}(t)a_2^{in}(t')^{\dagger}\rangle  = \delta(t-t'), \\
&& \langle f^{in}(t)f^{in}(t')^{\dagger}\rangle = \delta(t-t'), \\
&& \langle b^{in}(t)b^{in}(t')^{\dagger}\rangle = \left(\bar{N}+1\right)\delta(t-t'), \\
&& \langle b^{in}(t)^{\dagger} b^{in}(t')\rangle =\bar{N}
\delta(t-t'), \end{eqnarray} where $\bar{N}$ is mean thermal
vibrational number of the effective thermal reservoir coupling to
the atom center-of-mass motion~\cite{Footnote2}.

We assume that the laser is red-detuned and far-off resonance from
the atomic transition, i.e., $\Delta$ is negative and $|\Delta|$
is much larger than all the other parameters. This allows us to
eliminate adiabatically the atomic internal degrees of freedom,
and to assume that the atom always remains in the ground state $|g
\rangle $, that is, $\sigma_z(t) \approx -1$. Therefore we neglect
the time evolution of $\sigma_z$, Eq.~(\ref{lastqle}), while
Eq.~(\ref{penlastqle}) becomes
\begin{equation}
\dot{\sigma}(t)=-\left(\frac{\gamma}{2}-i\Delta \right)\sigma(t)-i
B(t)- \sqrt{\gamma}f^{in}(t), \label{lastqle2}
\end{equation}
whose formal solution is
\begin{eqnarray}
&& \sigma(t)=e^{-\left(\frac{\gamma}{2}-i\Delta \right)t}\sigma(0)  \label{lastqlesol} \\
&&-\int_0^t {\rm ds} e^{-\left(\frac{\gamma}{2}-i\Delta \right)s}
\left[i B(t-s)+ \sqrt{\gamma}f^{in}(t-s)\right]. \nonumber
\end{eqnarray} We now insert solution~(\ref{lastqlesol}) into the
other QLE and neglect the transient term because we are interested
in the dynamics at times which are much larger than $1/|\Delta|$.
We obtain \begin{widetext} \begin{eqnarray}
\dot{a}_1(t)&=&i\delta_1 a_1(t)+\int_0^t {\rm ds}
e^{-\left(\frac{\gamma}{2}-i\Delta \right)s}
\left[B(t-s)-i\sqrt{\gamma}f^{in}(t-s)\right]\left[B(t)^{\dagger}, a_1(t)\right] -\kappa_1 a_1(t)+\sqrt{2\kappa_1}a_1^{in}(t), \\
\dot{a}_2(t)&=&i\delta_2 a_1(t)+\int_0^t {\rm ds}
e^{-\left(\frac{\gamma}{2}-i\Delta \right)s}
\left[B(t-s)-i\sqrt{\gamma}f^{in}(t-s)\right]\left[B(t)^{\dagger}, a_2(t)\right] -\kappa_2 a_2(t)+\sqrt{2\kappa_2}a_2^{in}(t), \\
\dot{b}(t)&=&-i\nu b(t)+\int_0^t {\rm ds}
e^{-\left(\frac{\gamma}{2}-i\Delta \right)s}
\left[B(t-s)-i \sqrt{\gamma}f^{in}(t-s)\right]\left[B(t)^{\dagger}, b(t)\right] \nonumber \\
&&-\int_0^t {\rm ds} e^{-\left(\frac{\gamma}{2}+i\Delta \right)s}
\left[B(t-s)^{\dagger}+
i\sqrt{\gamma}f^{in}(t-s)^{\dagger}\right]\left[B(t), b(t)\right]
-\kappa_b b(t)+\sqrt{2\kappa_b}b^{in}(t), \end{eqnarray}
\end{widetext} where we have not taken care of operator ordering,
since, as we shall see, within the validity limit of our treatment
these integral terms will generate only linear contributions.

At this point, we choose the laser angular frequency $\omega_L$ so
that
\begin{eqnarray*} \delta_1 =\nu^{\prime}~~;~~\delta_2 =
-\nu^{\prime} \end{eqnarray*} namely, the laser frequency is tuned
symmetrically between the mode frequencies, which are spaced by a
quantity $2\nu^{\prime}$. The angular frequency
$\nu^{\prime}\simeq\nu$, and takes into account the a.c.-Stark
shifts due to the mechanical coupling with laser and cavity modes,
see Sec.~\ref{Sec:ac-Shift}, so that the two cavity modes are
resonant with the motional sidebands of the laser light. Together
with this choice of the laser frequency, we assume that the motional
sidebands are well resolved, that is, $\nu \gg |g_j|,\Omega,
\kappa_j$.

In order to identify the resonant process, we move to a frame
rotating at the effective vibrational angular frequency $\nu' \simeq
\nu$, (which has to be determined by solving the QLE) and we will
neglect in the QLE all the terms oscillating at $\nu'$ or larger.
Denoting the slowly varying quantities by
$\tilde{a}_1^{\dagger}(t)\equiv e^{i\nu 't} a_1^{\dagger}(t)$,
$\tilde{a}_2(t)\equiv e^{i\nu 't} a_2(t)$, $\tilde{b}(t)\equiv
e^{i\nu 't} b(t)$, after explicitly evaluating the commutators we
obtain \begin{widetext}
\begin{eqnarray} &&
\label{QLE_01} \dot{\tilde{a}}_1^{\dagger}(t)=i\left(\nu'-\delta_1
\right) \tilde{a}_1^{\dagger}(t) +\int_0^t {\rm ds}
e^{-\left(\frac{\gamma}{2}+i\Delta \right)s}
\left[B(t-s)^{\dagger}e^{i\nu 't}+i\sqrt{\gamma}f^{in}(t-s)^{\dagger}e^{i\nu 't}\right] \\
&& \times \left[-g_1 \cos \phi_1 \left(1-\frac{\eta^2}{2}\cos^2
\theta_c (2 \tilde{b}^{\dagger}\tilde{b}+1)\right)+\eta g_1
\sin\phi_1 \cos\theta_c\left(\tilde{b}(t)e^{-i\nu
't}+\tilde{b}^{\dagger}(t)e^{i\nu 't}\right)\right]
-\kappa_1 \tilde{a}_1^{\dagger}(t)+\sqrt{2\kappa_1}\tilde{a}_1^{in}(t)^{\dagger}, \nonumber \\
&& \dot{\tilde{a}}_2(t)=i\left(\nu'+\delta_2 \right)
\tilde{a}_2(t) +\int_0^t {\rm ds}
e^{-\left(\frac{\gamma}{2}-i\Delta \right)s}
\left[B(t-s)e^{i\nu 't}-i\sqrt{\gamma}f^{in}(t-s)e^{i\nu 't}\right] \\
&& \times \left[-g_2^* \cos \phi_2 \left(1-\frac{\eta^2}{2}\cos^2
\theta_c (2 \tilde{b}^{\dagger}\tilde{b}+1)\right) +\eta g_2^*
\sin\phi_2 \cos\theta_c\left(\tilde{b}(t)e^{-i\nu
't}+\tilde{b}^{\dagger}(t)e^{i\nu 't}\right)\right]
-\kappa_2 \tilde{a}_2(t)+\sqrt{2\kappa_2}\tilde{a}_2^{in}(t), \nonumber \\
&& \dot{\tilde{b}}(t)=i\left(\nu'-\nu \right) \tilde{b}(t)
+\int_0^t {\rm ds} e^{-\left(\frac{\gamma}{2}-i\Delta \right)s}
\left[B(t-s)e^{i\nu 't}-i\sqrt{\gamma}f^{in}(t-s)e^{i\nu 't}\right] \\
&& \times \left[i\eta \Omega^* \cos \theta_L +\eta g_1^*
\sin\phi_1 \cos\theta_c\tilde{a}_1^{\dagger}(t)e^{-i\nu 't}+
\eta g_2^* \sin\phi_2 \cos\theta_c\tilde{a}_2^{\dagger}(t)e^{i\nu 't}\right] \nonumber \\
&& -\int_0^t {\rm ds} e^{-\left(\frac{\gamma}{2}+i\Delta \right)s}
\left[B(t-s)^{\dagger}e^{i\nu 't}+i\sqrt{\gamma}f^{in}(t-s)^{\dagger}e^{i\nu 't}\right] \\
&& \times \left[-i\eta \Omega \cos \theta_L +\eta g_1 \sin\phi_1
\cos\theta_c\tilde{a}_1(t)e^{i\nu 't}+ \eta g_2 \sin\phi_2
\cos\theta_c\tilde{a}_2(t)e^{-i\nu 't}\right] -\kappa_b
\tilde{b}(t)+\sqrt{2\kappa_b}\tilde{b}^{in}(t), \label{QLE_0N}
\end{eqnarray}
\end{widetext} where we have introduced the noise operators
$\tilde{a}_1^{in}(t)\equiv e^{-i\nu 't} a_1^{in}(t)$,
$\tilde{a}_2^{in}(t)\equiv e^{i\nu 't} a_2^{in}(t)$, and
$\tilde{b}^{in}(t)\equiv e^{i\nu 't} b^{in}(t)$, which are still
delta-correlated.

We insert in these equations the explicit expression for $B(t-s)$,
thereby neglecting the terms oscillating at $\nu '$ or faster. We
finally perform the time integrals by making the Markovian
approximation $\exp\{-(\gamma/2\pm i\Delta +i m \nu')s\} \approx
\delta(s)/(\gamma/2\pm i\Delta +i m \nu')$, for $m=-1,0,1$. After
long, but straightforward calculations we get the final, effective
QLE at leading order in the Lamb-Dicke parameter, which read
\begin{widetext} \begin{eqnarray} \label{QLE:a1}
\dot{\tilde{a}}_1^{\dagger}(t)&=&i\left(\nu'-\delta_1 \right)
\tilde{a}_1^{\dagger}(t) +\chi_1^*\tilde{b}(t) -\left(\kappa_1
+\kappa_{1L}-i\delta_{1L}\right)
\tilde{a}_1^{\dagger}(t)+\sqrt{2\kappa_1}\tilde{a}_1^{in}(t)^{\dagger}
+\sqrt{2}\bar{\kappa}_{1L}\tilde{a}_{1L}^{in}(t)^{\dagger}+F_1,  \\
\label{QLE:a2} \dot{\tilde{a}}_2(t)&=&i\left(\nu'+\delta_2 \right)
\tilde{a}_2(t) +\chi_2\tilde{b}(t) -\left(\kappa_2
+\kappa_{2L}+i\delta_{2L}\right)
\tilde{a}_2(t)+\sqrt{2\kappa_2}\tilde{a}_2^{in}(t)
+\sqrt{2}\bar{\kappa}_{2L}\tilde{a}_{2L}^{in}(t)+F_2,  \\
\label{QLE:b} \dot{\tilde{b}}(t)&=&i\left(\nu'-\nu \right)
\tilde{b}(t) +\bar{\chi}_1 \tilde{a}_1^{\dagger}(t)-\bar{\chi}_2^*
\tilde{a}_2(t)
-\left(\kappa_b+\kappa_{2b}-\kappa_{1b}+i\delta_b\right)
\tilde{b}(t)\\
& &+\sqrt{2\kappa_b}\tilde{b}^{in}(t)
+\sqrt{2}\bar{\kappa}_{2b}\tilde{a}_{2L}^{in}(t)-\sqrt{2}\bar{\kappa}_{1b}\tilde{a}_{1L}^{in}(t)^{\dagger}+F_b~.\nonumber
\end{eqnarray} \end{widetext} Let us now discuss each term
appearing in the equations. The coupling coefficients are given by
\begin{eqnarray} &&\chi_1=\eta \Omega
g_1^*\cos\phi_1\left(\frac{\cos\theta_L}{\Delta-\nu'+{\rm
i}\gamma/2} +\frac{{\rm
i}\tan\phi_1\cos\theta_c}{\Delta+{\rm i}\gamma/2}\right)~,\nonumber\\
&&\label{Chi:1}\\
&&\chi_2=\eta \Omega
g_2^*\cos\phi_2\left(\frac{\cos\theta_L}{\Delta+\nu'+{\rm
i}\gamma/2}
+\frac{{\rm i}\tan\phi_2\cos\theta_c}{\Delta+{\rm i}\gamma/2}\right)~,\nonumber\\
&&\label{Chi:2} \\
&&\bar{\chi}_1=\eta \Omega
g_1^*\cos\phi_1\left(\frac{\cos\theta_L}{\Delta-\nu'-{\rm
i}\gamma/2} +\frac{{\rm
i}\tan\phi_1\cos\theta_c}{\Delta+{\rm i}\gamma/2}\right)~,\nonumber\\
&&\label{Chi:1_t}\\
&&\bar{\chi}_2=\eta \Omega
g_2^*\cos\phi_2\left(\frac{\cos\theta_L}{\Delta+\nu'-{\rm
i}\gamma/2}
+\frac{{\rm i}\tan\phi_2\cos\theta_c}{\Delta+{\rm i}\gamma/2}\right)~,\nonumber\\
&&\label{Chi:2:t}
\end{eqnarray}
and correspond to the Raman processes, in which laser photons are
scattered into the cavity mode with a change in the center-of-mass
excitation, see Sec.~\ref{Sec:Chi}.

New fluctuation-dissipation sources appear in the equations. We
first discuss noise terms appearing in Eqs.~(\ref{QLE:a1})
and~(\ref{QLE:a2}). In addition to cavity decay with rates
$\kappa_j$ we find processes described by the decay terms with
rate $\kappa_{1L}$ and $\kappa_{2L}$, and the corresponding
Langevin noises $\tilde{a}_{1L}^{in}(t)$ and
$\tilde{a}_{2L}^{in}(t)$, where
\begin{eqnarray}
&&\kappa_{1L}=\frac{\gamma}{2}\frac{|g_1|^2\cos^2\phi_1}{\gamma^2/4+(\Delta-\nu')^2}, \label{los1} \\
&&\kappa_{2L}=\frac{\gamma}{2}\frac{|g_2|^2\cos^2\phi_2}{\gamma^2/4+(\Delta+\nu')^2},
\label{los2}
\end{eqnarray}
and
\begin{eqnarray}
&&\tilde{a}_{1L}^{in}(t)=f^{in}(t)e^{-i\nu 't}, \label{nos1} \\
&&\tilde{a}_{2L}^{in}(t)=f^{in}(t)e^{i\nu 't}. \label{nos2}
\end{eqnarray}
These noises describe input-output processes between the cavity
modes and external modes, mediated by the atom. They possess the
same correlation functions of the spontaneous emission noise
$f^{in}(t)$, and at the timescales of interest, $\nu't \gg 1$,
they are uncorrelated from each other, thanks to the oscillating
factors. Note that
\begin{eqnarray}
&&\bar{\kappa}_{1L}=-{\rm i}\sqrt{\frac{\gamma}{2}}\frac{g_1\cos\phi_1}{\gamma/2+{\rm i}(\Delta-\nu')},\\
&&\bar{\kappa}_{2L}={\rm
i}\sqrt{\frac{\gamma}{2}}\frac{g_2^*\cos\phi_2}{\gamma/2-{\rm
i}(\Delta+\nu')}, \end{eqnarray} with
$\kappa_{jL}=|\bar{\kappa}_{jL}|^2$. They originate from the
scattering processes in which cavity photons are lost because they
are absorbed and then spontaneously emitted by the atom, as has
been discussed in Sec.~\ref{Sec:cav-noise}.

The noise and dissipation terms in Eq.~(\ref{QLE:b}), in addition
to the noise terms of the trap, are described by the decay terms
with rate $\kappa_{1b}$ and $\kappa_{2b}$, and the corresponding
Langevin noise operators $\tilde{a}_{1L}^{in}(t)$ and
$\tilde{a}_{2L}^{in}(t)$. These processes originate from
incoherent emission or absorption of a vibrational quantum
accompanied by absorption and subsequent spontaneous emission of a
laser photon. The emission of vibrational quanta takes place at
rate \begin{equation}\label{absphon2}
\kappa_{2b}=\frac{\gamma}{2}\frac{\eta^2|\Omega|^2\cos^2\theta_L}{\gamma^2/4+(\Delta+\nu')^2},
\label{losphon1} \end{equation} while the rate of incoherent
absorption of vibrational quanta is given by
\begin{equation}\label{absphon1}
\kappa_{1b}=\frac{\gamma}{2}\frac{\eta^2|\Omega|^2\cos^2\theta_L}{\gamma^2/4+(\Delta-\nu')^2}.
\label{losphon2} \end{equation} In particular, when $\Delta <0$
then $\kappa_{2b} > \kappa_{1b}$ and the motion is cooled.
Moreover, \begin{eqnarray}
&&\bar{\kappa}_{1b}=\sqrt{\frac{\gamma}{2}}\frac{\eta\Omega\cos\theta_L}{\gamma/2+{\rm i}(\Delta-\nu')},\\
&&\bar{\kappa}_{2b}=\sqrt{\frac{\gamma}{2}}\frac{\eta\Omega^*\cos\theta_L}{\gamma/2-{\rm
i}(\Delta+\nu')},
\end{eqnarray}
with $\kappa_{jb}=|\bar{\kappa}_{jb}|^2$. If we consider the
dynamics described by these terms only, these incoherent phonon
absorption and emission processes lead to thermalization of the
atomic motion at rate $\kappa_{2b}-\kappa_{1b}$, to a final
effective mean vibrational number
$n_{th}=\kappa_{1b}/(\kappa_{2b}-\kappa_{1b})\simeq |\Delta
|/4\nu'$, as in standard cooling~\cite{Stenholm86}. However, the noise associated
with these incoherent phonon absorptions and emissions is
\emph{correlated} with the noise terms $\tilde{a}_{1L}^{in}(t)$
and $\tilde{a}_{2L}^{in}(t)$ describing scattering of cavity
photons, because all these processes ultimately originate from
spontaneous emission. This is why the noise terms in the Langevin
equation for the atomic motion are directly expressed in terms of
$\tilde{a}_{1L}^{in}(t)$ and $\tilde{a}_{2L}^{in}(t)$, making
therefore this correlation evident.

The operators $F_j$ in Eqs.~(\ref{QLE_01})-(\ref{QLE_0N})
represent non-linear terms, which describe the noise associated
with the incoherent part of the scattering processes discussed in
Sec.~\ref{Sec:Chi}. These terms can be neglected with respect to
the coherent processes, provided that $\gamma\ll|\Delta|$ and
$\gamma\ll\nu$. In particular, the second inequality ensures that
rates $\chi_1$ and $\chi_2$ differ appreciably, such that
entanglement between the cavity modes can be established in a
finite time~\cite{Morigi06}. We will focus on this regime,
$\gamma\ll\nu$, in which we can thus neglect $F_j$ in the
effective QLE when evaluating the spectrum of squeezing.

Finally, the frequency shifts of the two cavity modes and of the
vibrational motion read
\begin{eqnarray}
&&\delta_{1L}=\frac{(\Delta-\nu')|g_1|^2\cos^2\phi_1}{\gamma^2/4+(\Delta-\nu')^2} \label{eqfornu1} \\
&&\delta_{2L}=\frac{(\Delta+\nu')|g_2|^2\cos^2\phi_2}{\gamma^2/4+(\Delta+\nu')^2} \label{eqfornu2} \\
&&\delta_{b}= \frac{2\Delta \eta^2 |\Omega|^2
\cos^2\theta_L\left(\gamma^2/4+\Delta^2-\nu'^2\right)}{\left(\gamma^2/4+\Delta^2-\nu'^2\right)^2
+\nu'^2\gamma^2}\label{nu_b} \label{eqfornu}\\
&&~-\eta^2 |\Omega|^2
\cos^2\theta_L\frac{\Delta}{\Delta^2+\gamma^2/4}\nonumber
\end{eqnarray}
and from their form one can recognize the a.c.-Stark shifts
reported in Sec.~\ref{Sec:ac-Shift}, with
$\delta\omega_j=\delta_{jL}a^{\dagger}_ja_j$,
Eq.~(\ref{deltaomega_j}), and $\delta\nu_b=\delta_bb^{\dagger}b$,
Eq.~(\ref{deltanu_b}), where now $\nu\to\nu^{\prime}$.
Note that we have omitted a non-linear shift at
second order in the Lamb-Dicke parameter, which affects both
cavity modes and motion. As discussed in Sec.~\ref{Sec:ac-Shift},
this is a small correction to $\delta_{jL}$, as it scales with
$\eta^2$, while it may have a relevant effect on the
center-of-mass dynamics. It can be neglected in the limit
$\Omega\cos^2\theta_L\gg g_j\cos^2\theta_c$. Under this
assumption, which we will consider in the rest of this manuscript,
the spectrum of the center-of-mass is the spectrum of a harmonic oscillator,
characterized by equidistant energy levels.

As the dynamics we seek relies on resonant interaction between the
cavity modes and the vibrational motion, the two cavity modes should
be exactly at resonance with the sidebands of the driving laser.
Equation~(\ref{nu_b}) provides an implicit equation for the actual
vibrational angular frequency $\nu'$. In the parameter regime $\eta
|\Omega | \ll |\Delta|, \nu$ we find with good approximation
\begin{eqnarray}\label{renfreq}
  \nu'&\approx& \nu+\frac{2\Delta \eta^2 |\Omega|^2
  \cos^2\theta_L\left(\gamma^2/4+\Delta^2-\nu^2\right)}{\left(\gamma^2/4+\Delta^2-\nu^2\right)^2+\nu^2\gamma^2}\nonumber\\
&   &~-\eta^2 |\Omega|^2
\cos^2\theta_L\frac{\Delta}{\Delta^2+\gamma^2/4}.
\end{eqnarray}
Taking also into account the frequency shifts of
Eqs.~(\ref{eqfornu1})-(\ref{eqfornu2}), the resonance conditions
are finally
\begin{eqnarray}
&&\delta_1=\delta_{1L}+\nu' \label{res1} \\
&&\delta_2=\delta_{2L}-\nu' \label{res2} .
\end{eqnarray}

In the parameter regime $\gamma\ll\nu$, using conditions
(\ref{res1})-(\ref{res2}), we arrive therefore to the final QLE,
describing the coherent interaction between the two cavity modes
and the vibrational motion, competing with losses and noise
processes due to spontaneous emission, cavity decay, and
vibrational heating,
\begin{widetext} \begin{eqnarray} &&
\dot{\tilde{a}}_1^{\dagger}(t)=\chi_1^*\tilde{b}(t)
-\left(\kappa_1 +\kappa_{1L}\right)
\tilde{a}_1^{\dagger}(t)+\sqrt{2\kappa_1}\tilde{a}_1^{in}(t)^{\dagger}
+\sqrt{2}\bar{\kappa}_{1L}\tilde{a}_{1L}^{in}(t)^{\dagger}~,  \label{qlefin1}\\
&& \dot{\tilde{a}}_2(t)=\chi_2\tilde{b}(t) -\left(\kappa_2
+\kappa_{2L}\right)
\tilde{a}_2(t)+\sqrt{2\kappa_2}\tilde{a}_2^{in}(t)
+\sqrt{2}\bar{\kappa}_{2L}\tilde{a}_{2L}^{in}(t)~,  \label{qlefin2} \\
&& \dot{\tilde{b}}(t)=\bar{\chi}_1
\tilde{a}_1^{\dagger}(t)-\bar{\chi}_2^* \tilde{a}_2(t)
-\left(\kappa_b+\kappa_{2b}-\kappa_{1b}\right)
\tilde{b}(t)+\sqrt{2\kappa_b}\tilde{b}^{in}(t)
+\sqrt{2}\bar{\kappa}_{2b}\tilde{a}_{2L}^{in}(t)-\sqrt{2}\bar{\kappa}_{1b}\tilde{a}_{1L}^{in}(t)^{\dagger}~.
\label{qlefin3}
\end{eqnarray}
\end{widetext}

\subsection{Evaluation of the spectrum of squeezing}

We now use Eqs.~(\ref{qlefin1})-(\ref{qlefin3}) in order to
determine the stationary spectrum of squeezing of the light at the
cavity output. We consider
\begin{eqnarray}\label{imeno}
I_-^{out}(t)&=&
a_1^{out}(t)+a_1^{out}(t)^{\dagger}-a_2^{out}(t)-a_2^{out}(t)^{\dagger},
\\
I_+^{out}(t)&=&-i\left[
a_1^{out}(t)-a_1^{out}(t)^{\dagger}+a_2^{out}(t)-a_2^{out}(t)^{\dagger}\right],\label{ipiu}
\end{eqnarray} corresponding respectively to the difference
between the amplitude quadratures, and the sum of the phase
quadratures of the two sideband modes. These are the quadratures
exhibiting two-mode squeezing in the case of pulsed excitation in
this setup, see \cite{PRL,Morigi06}. The output cavity fields
$a_j^{out}(t)$ in Eqs.~(\ref{imeno})-(\ref{ipiu}) are given by the
usual input-output relation \begin{equation} \label{aout}
a_j^{out}(t)=\sqrt{2\kappa_j}a_j(t)-a_j^{in}(t), \;\;\;\;j=1,2.
\end{equation} The spectrum of squeezing can be calculated by
evaluating the Fourier transforms \begin{equation}
\hat{I}_{\pm}^{out}(\omega)=\int{\rm d}t{\rm e}^{{\rm i}\omega
t}I_{\pm}^{out}(t), \end{equation} and using the fact that at the
stationary state it is \begin{equation}\label{defspec} \langle
\hat{I}_{\pm}^{out}(\omega) \hat{I}_{\pm}^{out}(\omega')+
\hat{I}_{\pm}^{out}(\omega') \hat{I}_{\pm}^{out}(\omega)\rangle =
8\pi S_{\pm}(\omega)\delta(\omega+\omega'), \end{equation} where
we have normalized the spectrum so that the shot noise level
corresponds to $S_{\pm}(\omega)=1$. Two-mode squeezing is found
when one spectrum of squeezing takes values below the shot noise
limit at some $\omega$. From the Fourier transform of
Eqs.~(\ref{qlefin1})-(\ref{qlefin3}) one can see that
$S_{+}(\omega)=S_{-}(\omega)\equiv S(\omega)$, which implies that
in the present case two-mode squeezing is equivalent to EPR-like
entanglement between the two output cavity modes. This is easily
verified by applying a sufficient criterion for entanglement, such
as the ``sum'' criterion of Duan \textit{et al.} \cite{Duan00}, or
the product criterion of Ref.~\cite{Mancini02,GIOV03}. With the
chosen normalization for the output cavity modes at $\omega$, the
sum criterion reads \begin{equation} S_{+}(\omega)+S_{-}(\omega) <
2, \label{duan} \end{equation} while the product criterion gives
\begin{equation} S_{+}(\omega)S_{-}(\omega) < 1, \label{noi}
\end{equation} so that in our case both criteria imply that the
two output modes are EPR-like entangled as soon as $S(\omega) <
1$. The squeezing spectrum $S(\omega)$ can be obtained from the
Fourier transform of the Langevin equations after long but
straightforward algebra, yielding a cumbersome expression which
will not be reported here. This expression becomes considerably
simpler in the limit $|\Omega|, |g_j|, \gamma \ll |\Delta|$ and
$\eta \ll 1$. In this limit the additional loss processes due to
spontaneous emission, associated with the rates $\kappa_{jL}$ and
$\kappa_{jb}$ ($j=1,2$), are typically negligible,
that is, $\kappa_{jL},\kappa_{jb} \ll\kappa$.
Moreover, we consider
the case of ion traps, where heating of the atomic motion is
negligible with respect to all radiative noise
sources~\cite{Trap:stability}. Finally, as the two cavity modes are
very close in frequency, they will have very similar properties, in
particular we can take $\kappa_1=\kappa_2=\kappa$. In this parameter
regime the main aspects of the squeezing spectrum can be grasped
from its analytical expression. One finds
\begin{equation}\label{specsimple}
S(\omega)=1-\frac{\kappa^2\left(\Theta^4-\Sigma^4\right)}{\left(\kappa^2+\omega^2\right)
\left[\left(\omega^2-\Theta^2\right)^2+\omega^2\kappa^2\right]},
\end{equation}
where $\Theta=\sqrt{|\chi_2|^2-|\chi _1|^2}$ as given in
Eq.~(\ref{thetabig}), and
\begin{equation} \Sigma =
\sqrt{\left|\left|\chi_2\right|^2+\left|\chi_1\right|^2-2\chi_1\chi_2\right|},
\end{equation}
and we have used that $\chi_j=\bar{\chi}_j$ when $\gamma \ll
|\Delta|$ (see Eqs.~(\ref{Chi:1})-(\ref{Chi:2:t})).
Note that due to the transformations which we have
applied, the results which appear around $\omega=0$ in $S(\omega)$
describe quantum correlations of noise components in the optical
signal at $\omega_L - \nu' \pm \omega$ with those at $\omega_L +
\nu' \pm \omega$, i.e.\ correlated fluctuations of the two modes
at the same offset from their center frequencies.

From Eq.~(\ref{specsimple}) one notes that the properties of the
spectrum are mainly determined by the ratio $\Theta/\kappa$. In
fact, the denominator in Eq.~(\ref{specsimple}) has always three
poles in the lower complex half-plane, one which is always imaginary
at $\omega_0=-i\kappa$, and two poles at $\omega_{\pm}=-i\kappa/2\pm
\sqrt{\Theta^2-\kappa^2/4}$. Therefore, when $\Theta/\kappa \gg 1$
the two poles at $\omega_{\pm}$ have a nonzero real part and the
spectrum is characterized by three well-separated inverted
Lorentzian peaks, one at $\omega=0$ with width (FWHM) $2\kappa$ and
the other two symmetrically placed at $\omega\approx \pm \Theta$,
with FWHM $\kappa$. At the center of these peaks one has
$S=(\Sigma/\Theta)^4 \simeq
\left[\left(|\chi_2|-|\chi_1|\right)/\left(|\chi_2|+|\chi_1|\right)\right]^2$,
approaching $S(\omega)=0$, i.e., infinite two-mode squeezing, for
$|\chi_2|\simeq |\chi_1|$. Therefore, when $\Theta>\kappa$ we find
two-mode squeezing within three narrow bandwidths around $\omega=0$
and $\omega =\pm \Theta$. In the opposite case of $\Theta \leq
\kappa/2$, the three poles are all on the imaginary axis, and the
spectrum shows only one inverted Lorentzian peak at $\omega=0$. When
$\Theta/\kappa \ll 1$, this peak becomes very narrow, with FWHM
$\sim 2\Theta^2/\kappa$.

It is remarkable that even for $\kappa>\Theta$ one finds almost
perfect squeezing in the difference of amplitude quadratures at
$\omega=0$. This can be understood, as in the regime we consider
the scattered photons due to spontaneous emission are negligible
with respect to those lost through the output cavity mirror
($\kappa_{jL},\kappa_{jb} \ll \kappa)$. This implies that most of
the intracavity photons are detected at the output. These photons
are almost perfectly correlated at $\omega=0$ and therefore would
give $S(0) \simeq 0$. In this regime, a large cavity decay rate
$\kappa$ has only the effect of narrowing the squeezing bandwidth.
On the contrary, if the photon scattering by spontaneous emission
is no more negligible, two-mode squeezing soon degrades, even at
$\omega=0$ (see for example \cite{milwal}).

The presence of the three-pole structure in the squeezing spectrum
is novel with respect to the spectral features usually encountered
in the parametric oscillator (either below and above threshold,
see e.g.\ Ref.~\cite{milwal}). This structure is due to the
coherent interaction of the two cavity modes with the quantized
atomic motion, i.e.\ it arises from the coherent microscopic
processes underlying the dynamic establishment of quantum
correlations. The peculiar spectral properties can be exploited to
achieve \emph{optimal broadband two-mode squeezing} when the three
peaks merge, which happens when $\Theta =\kappa$. In this case one
easily sees from Eq.~(\ref{specsimple}) that
\begin{equation}\label{specsimpleflat}
S(\omega)=1-\frac{\kappa^2\left(\Theta^4-\Sigma^4\right)}{\Theta^6
+\omega^6}, \end{equation}
i.e., one has large, uniform squeezing for a wide bandwidth of
frequencies. The fact that $\Theta = \kappa$ is the condition for
the best two-mode squeezing in the output can be easily understood
noticing that $\Theta$ is the angular frequency at which two-mode
squeezing inside the cavity is periodically built up (see
Ref.~\cite{PRL, Morigi06}). Therefore, when $\Theta =\kappa$,
squeezing is generated inside the cavity at the same rate at which
it is transferred to the output field. In contrast, in the other two
cases, squeezing is not efficiently generated in the output because
either the output coupling happens too fast, i.e.\ before full
intra-cavity squeezing is established, or the output coupling is
sufficiently slow to allow that energy is stored inside the cavity
and the two-mode squeezing is coherently re-converted into
independent states before it is coupled out.

\subsection{Results} \label{Sec:Results}

We now consider the exact squeezing spectrum $S(\omega)$, defined
in Eq.~(\ref{defspec}), where the operator~(\ref{imeno}) is
evaluated using the output relation~(\ref{aout}), such that the
operators $a_j(t)$, $a_j^{\dagger}(t)$ are the solutions of the
QLE~(\ref{qlefin1})-(\ref{qlefin3}) including all noise and loss
terms.

\begin{center} \begin{figure}[htb]
\epsfig{width=0.45\textwidth, file=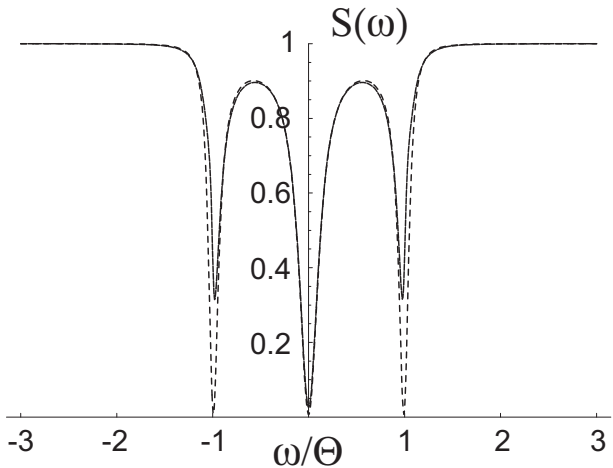} %
\caption{Squeezing spectrum $S(\omega)$ as a function of the
sideband frequency $\omega$ (in units of $\Theta$) when $\Theta
\simeq 8 \kappa$. Its behavior is well reproduced by the
approximate analytical expression of
Eq.~(\protect\ref{specsimple}) (dashed line). The parameters are
discussed in the text. } \label{Fig:2}
\end{figure}
\end{center}

The parameter regime we consider has been discussed in detail
in~\cite{Morigi06}. We take a ${F}\!=0 \leftrightarrow
{F^{\prime}}\!=\!1$ atomic transition with the quantization axis
$\vec{B}$ along the cavity axis, and $\vec{B}$, $\vec{k}_L$, and
laser polarization $\vec{E}_L$ mutually orthogonal. Relation
$\gamma\ll\nu$ can be fulfilled by the intercombination line of an
Indium ion at $\gamma=2\pi\times 360$~kHz in a trap at
$\nu=2\pi\times 3$~MHz, for which $\eta \simeq 0.1$. We consider a
geometrical configuration corresponding to $\theta_L=0$,
$\theta_c=\pi/2$, and $\phi_1=\phi_2=0$, which means that the ion
motion takes place along the direction of the laser beam and
orthogonally to the cavity axis, and that the trap center
coincides with an antinode of the cavity modes. In such a case,
the coupling coefficients of Eqs.~(\ref{Chi:1})-(\ref{Chi:2:t})
are determined by the first term only. Moreover in this case 
$\Omega\cos^2\theta_L\gg g_j\cos^2\theta_c \simeq 0$, and
therefore, as discussed in Secs.~\ref{Sec:ac-Shift} and
\ref{Sec:QLE}, the small Stark-shift correction to $\delta_{jL}$,
scaling with $\eta^2$ can be neglected. If we consider that the
ion couples to two non-degenerate polarization modes of a
resonator with vacuum Rabi couplings $g\approx 2\pi\times
0.6$~MHz, and we take laser Rabi frequency $\Omega=2\pi\times
18$~MHz and detuning $\Delta=2\pi\times 60$~MHz, we obtain
$\Theta/2\pi \simeq 7.9$~kHz, see~\cite{Morigi06}. The condition
$\Theta \gg \kappa$ is found for a finesse ${\cal F}\simeq 10^6$
and free spectral range $\delta\omega= 2\pi\times 1$~GHz, so that
$\kappa=2\pi\times 1$~KHz. The corresponding spectrum of squeezing
is displayed in Fig.~\ref{Fig:2} (full line). It exhibits three
minima at $\omega=0,\pm\Theta$, which correspond to three
separated regions of narrow-band squeezing. The two bands around
$\omega = \pm\Theta$ have width $\kappa$, while the central one
has width $2 \kappa$ and shows almost $100\%$ squeezing. These
features are well reproduced by the analytical
expression~(\ref{specsimple}) (see dashed line in
Fig.~\ref{Fig:2}), except that the latter predicts very large
squeezing also for the peaks at $\omega = \pm\Theta$. The success
of the simplified expression~(\ref{specsimple}) is due to the fact
that, with the chosen parameter values, the loss rates due to the
various scattering processes are at least ten times smaller than
the cavity output loss rates $\kappa_1=\kappa_2=\kappa$, and
therefore do not have a relevant effect on the spectrum. We have
also considered a realistic ion vibrational heating rate
$\kappa_h=\kappa_b\bar{N}=2\pi \times 0.1$~kHz, which however
gives an effect which is negligible even with respect to that due
to photon scattering. The appearance of three minima is a novel
behavior to our knowledge, and it arises from the coherent
microscopic dynamics, as $\Theta$ modulates the exchange of
excitations and correlations between the cavity modes and the
center-of-mass motion.

\begin{center} \begin{figure}[htb]
\epsfig{width=0.45\textwidth, file=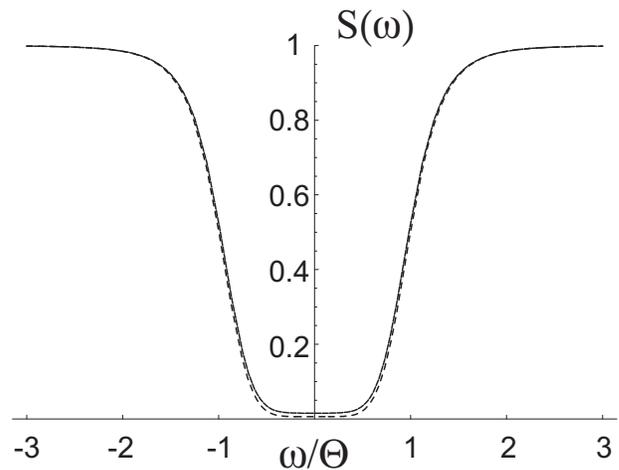} %
\caption{Squeezing spectrum $S(\omega)$ when $\Theta = \kappa$.
Its behavior is well reproduced by the approximate analytical
expression of Eq.~(\protect\ref{specsimple}) (dashed line). The
parameters are discussed in the text.} \label{Fig:3} \end{figure}
\end{center}

The most interesting regime of broadband two-mode squeezing, when
$\Theta\sim\kappa$, is shown in Fig.~\ref{Fig:3}, which displays
the squeezing spectrum in the case of the same parameter values of
Fig.~\ref{Fig:2} except for a lower cavity finesse, ${\cal
F}\simeq 10^5$, implying $\kappa=2\pi\times 10$~kHz. The three
minima merge into a single broad one, centered around $\omega=0$,
whose width is determined by $\kappa=\Theta$. Also in this case
one gets almost perfect squeezing at the center, and these
features are well reproduced by the simple analytical expression
of Eq.~(\ref{specsimpleflat}) (dashed line in Fig.~\ref{Fig:3}).

\begin{center} \begin{figure}[htb] \epsfig{width=0.45\textwidth,
file=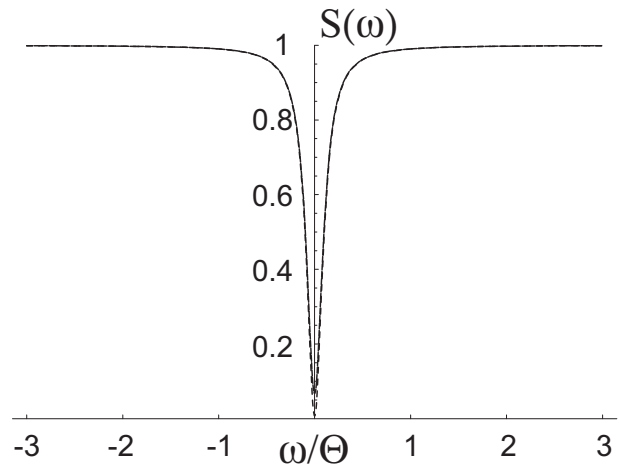} \caption{Squeezing spectrum $S(\omega)$
when $\Theta \simeq \kappa/10$. Its behavior is well reproduced by
the approximate analytical expression of
Eq.~(\protect\ref{specsimple}) (dashed line). The parameters are
discussed in the text.} \label{Fig:4} \end{figure} \end{center}

Finally, in Fig.~\ref{Fig:4} we consider the case $\kappa>\Theta$.
We have still kept the parameter values of Fig.~\ref{Fig:2}, but
we have now considered a cavity with finesse ${\cal F} \sim 10^4$,
implying $\kappa=2\pi\times 100$~kHz. The squeezing features are
visibly worsened, as in this regime losses are faster than the
typical time scale in which correlations between the field modes
are established. One has still two-mode squeezing around
$\omega=0$, but with a very narrow bandwidth which is roughly
given by $\Theta^2/2\kappa$.

\section{Conclusions}
\label{Sec:Conclusions}

The resonance fluorescence of a confined, single, laser-driven
atom exhibits EPR-entanglement, or two-mode squeezing, in the
field modes which interact resonantly with the Stokes and
anti-Stokes transitions created by the atomic motion. By coupling
these sidebands to a high-finesse optical cavity, we have shown
how to create continuous-wave (cw) two-mode squeezed light output
from the cavity. At the microscopic level, the process is based on
the mechanical effect of light, which allows for quantum-coherent
generation and control of entanglement between the motion and the
cavity modes. The scattering processes have been characterized and
described in simple physical pictures using scattering matrix
theory, and the squeezing spectrum has been calculated using
Quantum Langevin Equations.

Peculiar novel spectral properties are predicted for the squeezing
spectrum of the cavity output. They may be divided into three
regimes of the cavity output rate $\kappa$ relative to the
frequency $\Theta$ of creation of two-mode squeezing inside the
cavity. The squeezing spectrum can consist of a single peak
($\kappa > \Theta$), three peaks ($\kappa < \Theta$), or one
broad, homogeneous band ($\kappa = \Theta$). Simple analytical
approximations have been derived for the three relevant regimes.

The squeezing spectrum in the different parameter regimes has been
calculated for an experimentally accessible case of a single
trapped ion as a specific example. The results for this specific
system show all the features predicted by the general derivations,
exhibiting novel spectral properties of two-mode squeezing which
are novel when compared with conventional
Optical-parameter-amplifier-type of sources.

In particular, starting from the most fundamental individual
quantum systems, a single atom and an optical cavity, we have
designed a nonlinear optical source. This is therefore a
paradigmatic model system exhibiting the connection between
microscopic, quantum-coherent dynamics and macroscopic nonlinear
device properties. Its efficiency and the high-degree of control
one can achieve on its dynamics offer promising perspectives for
the realization of quantum light sources for quantum
networking~\cite{CiracKimble, Kraus04}.

\section{Acknowledgements}

The authors acknowledge stimulating discussions with and helpful
comments from P. Ca\~nizares, S. Mancini and P. Tombesi. One of
the authors (D.V.) acknowledges the Grup d'Optica at the
Universitat Autonoma de Barcelona for hospitality during
completion of this work. This work was partly supported by the
European Commission (CONQUEST network, MRTN-CT-2003-505089; SCALA,
Contract No.\ 015714) and by the Spanish Ministerio de Educaci\'on
y Ciencia (LACSMY, FIS2004-05830; Ramon-y-Cajal fellowship; QLIQS,
FIS2005-08257).

\end{document}